\documentclass[fleqn,twoside]{article}
\usepackage{espcrc2}
\usepackage{amssymb}
\usepackage{amsmath}

\usepackage{graphicx}
\usepackage[figuresright]{rotating}

\nopagebreak
\title{{
\vskip-20pt 
\hfill {\rm\normalsize FERMILAB-Conf-03/379-A}}\break
Neutrino Mixing and Cosmology}

\author{Nicole F. Bell\address{NASA/Fermilab Astrophysics Center, 
Fermi National Accelerator Laboratory, Batavia, Illinois 60510-0500}
\thanks{nfb@fnal.gov}
\thanks{Talk presented at TAUP 2003, Seattle, USA, 5-9 September 2003.}}

\begin{document}

\begin{abstract}

We review the consequences of neutrino mixing in the early universe.  For both
active-sterile mixing or mixing between three active neutrinos only, the
consequences of oscillations depend crucially upon the size of the universe's
lepton number (relic neutrino asymmetry.)
\vspace{1pc}
\end{abstract}

\maketitle

\section{Introduction}

The relic neutrino background has never been directly detected, so we must
resort to indirect means to infer its properties.  One of the most useful 
tools available is Big Bang Nucleosynthesis (BBN).
%
%
By putting neutrino oscillations together with BBN, we may shed light on
neutrino mixing, cosmology, or both.
We will outline the consequences of oscillations of the relic
background neutrinos, for both active-active and active-sterile oscillation 
modes.   The central issue we shall probe is:
how big is the universe's lepton number?

While the baryon asymmetry of the universe is well determined, 
$n_{B}/n_\gamma \simeq 5 \times 10^{-10}$, the size of the lepton asymmetry is
unknown.  The simplest assumption is that the baryon and lepton asymmetries
are of the same size, as would be the case if $B-L$ were conserved.  However,
there are viable models in which $L$ may be large while $B$ is small.  
Given constraints on charge neutrality, any large lepton asymmetry would have
to be hidden in the neutrino sector.

Since neutrinos and antineutrinos should be in chemical equilibrium
until they decouple at a temperature $T \sim 2$ MeV, they may be
well-described by Fermi-Dirac distributions with equal and opposite
chemical potentials:
\begin{equation}
f(p,\xi) = \frac{1}{1+\exp(p/T-\xi)}\,,
\end{equation}
where $p$ denotes the neutrino momentum, $T$ the temperature, and $\xi$
the chemical potential in units of $T$.  
The lepton
asymmetry $L_{\alpha}$ for a given flavour $\nu_{\alpha}$ is related to
the chemical potential by
\begin{equation}
L_{\alpha} = \frac{n_{\nu_\alpha} - n_{\bar{\nu}_\alpha}}{n_\gamma}
= \frac{\pi^2}{12\zeta(3)}\left(\xi_\alpha + 
\frac{\xi_\alpha^3}{\pi^2}\right)\,,
\label{leptxi}
\end{equation}
where $\zeta(3)\simeq 1.202$. 
A nonzero chemical potential results in extra energy density, such that the 
effective number of neutrinos is increased from the standard model prediction 
by
\begin{equation}
\Delta N_\nu = \frac{30}{7}\left(\frac{\xi}{\pi}\right)^2 +
\frac{15}{7}\left(\frac{\xi}{\pi}\right)^4.
\label{delnu}
\end{equation}

Large chemical potentials affect BBN in two ways:
\begin{enumerate}
\item
The extra energy density increases the expansion rate of the universe, thus
increasing the BBN helium abundance, and also alters the CMB results.
This sets the weak bound 
$|\xi_\alpha| \lesssim 3$, for all three flavours.

\item
An additional, much stronger, limit can be placed on the $\nu_e$ --
$\overline{\nu}_e$ asymmetry, as it directly effects the neutron to proton
ratio prior to BBN by altering beta-equilibrium.  
(Beta-equilibrium is between the weak interactions $n+\nu_e\leftrightarrow
p+e^-$ and $p+\bar\nu_e\leftrightarrow n+e^+$.)
For example, positive $\xi_e$ increases the $\nu_e$ abundance relative to
$\bar{\nu}_e$, thus lowering the neutron to proton ratio and decreasing the
helium yield.  This sets the limit $|\xi_e| \lesssim 0.04$.
\end{enumerate}
However, it is possible that the two effect compensate for each other, 
i.e., the effects of a small $\xi_e$ are partially undone by an increased
expansion rate due to a large $\xi_{\mu,\tau}$.  In this case the bounds 
become~\cite{oldbounds}:
\begin{eqnarray}
\label{oldlimits}
-0.01 < \xi_e & < & 0.22\,, \\
|\xi_{\mu,\tau}| &<& 2.6\,,
\end{eqnarray}
where the upper limits are obtained only in tandem.

\section{Oscillations between three active neutrino}

Since we know neutrinos oscillate, the individual lepton numbers 
$L_e$, $L_\mu$ and $L_\tau$ are violated and only the total lepton number 
is conserved. 
It was suggested in Ref.~\cite{smirnov} (see also \cite{fuller})
that the large neutrino mixing angles implied by the present data may lead 
to equilibration of all three flavours in the early universe. 
If a large  asymmetry hidden in $\xi_{\mu,\tau}$ were to be transfered to 
$\xi_e$ well before weak freezeout at $T\simeq 1\rm\ MeV$, the 
stringent BBN limit on $\xi_e$ would then apply to all three flavours, 
improving the bounds on $\xi_{\mu,\tau}$
by nearly two orders of magnitude.

This proposal was recently studied in detail in Refs.~\cite{dhpprs,abb,yyy}, 
where close to complete equilibration  of the asymmetries $\xi_{\mu,\tau}$
with $\xi_e$ was found.  
The equilibration takes place when $T \sim 2\rm\ MeV$ via an MSW 
transition which is driven  by the solar mass squared difference.
(It is actually more complicated than a normal 
MSW transition, as forward scattering from other neutrinos in the background
medium introduces a non-linear contribution to the effective 
potential~\cite{pantaleone}.
This has the effect of synchronizing the neutrino ensemble
such that all momentum modes go through the MSW resonance together, with
parameters governed by an effective momentum, which is close to the thermal
average.  See \cite{dhpprs,abb,yyy,prs} for details.)
This MSW transition converts the initial flavour
states (which are approximately mass eigenstates at high temperature) into
vacuum mass eigenstates, which, due to the large solar mixing angle, 
have large components of all three flavours.

This equilibration is an important
result, as it excludes the possibility of degenerate
BBN~\cite{oritoesposito}, and is the strongest limit on the total
lepton number of the universe and is likely to remain so for the
foreseeable future.
In terms of extra relativistic degrees of freedom, the limit is 
impressively tight: If $\xi_{e,\mu,\tau} = 0.04$, then 
$\Delta N_\nu \simeq 3 \times 0.0007 = 0.002$.
One implication is that cosmological constraints on (and future measurements 
of) neutrino masses will not be subject to uncertainty in the relic neutrino 
density.

Strictly speaking, one version of the degenerate BBN scenario is still
possible: It is conceivable that $\xi_e \sim \xi_{\mu} \sim \xi_{\tau} \sim
0.2$, provided that another relativistic particle species contributes the extra
energy density required to compensate for the large $\nu_e$ chemical
potential~\cite{abb,kneller}.  This extra energy density can no longer consist
of active neutrino, so would have to be something more exotic.  Such an
unnatural scenario could eventually be detected via the CMB.

\section{Active-sterile oscillations}

Active-sterile oscillations before the time of BBN are potentially dangerous,
as they may bring the sterile degrees of freedom into thermal equilibrium, thus
increasing the expansion rate and thereby upsetting the successful predictions
of BBN.  All models which seek to explain LSND~\cite{lsnd} via the addition of
a sterile neutrino suffer from this problem.  Since the LSND mixing angle is 
relatively large, the sterile neutrino is always thermalised~\cite{sterile}.

The rate at which the sterile species is populated is
\begin{equation}
\Gamma(\nu_\alpha \rightarrow \nu_s) \simeq \frac{1}{2} \Gamma_{\rm scatt} 
\sin^2 2\theta_m,
\end{equation} 
where $\Gamma_{\rm scatt}$ is he active-neutrino collision rate, and $\theta_m$
is the matter affected mixing angle.
However, it is possible to avoid thermalisation of the sterile, by invoking the
presence of a small relic neutrino asymmetry \cite{fv}.  This provides an index
of refraction, which suppresses the active-sterile mixing angle so that the
rate of sterile production is negligible, until after the neutrinos thermally
decouple from the rest of the plasma.

\section{Cosmological neutrino mass limits, and 4-neutrino models}

The absolute neutrino mass scale may be probed via cosmological means through 
large scale structure measurements, as
free streaming of neutrinos suppresses the growth of structure 
on the small scales that are within the horizon while the neutrinos are 
relativistic.
The current limit is roughly $\sum m_\nu \lesssim 1 {\rm eV}$, depending 
somewhat on how conservatively parameter degeneracies are priors are  
treated~\cite{hannestad}.

Since the LSND mass squared difference is $\sim O({\rm eV})$, if all 4
neutrino species are populated in the universe, the cosmological mass limits
constrain the LSND parameter space. However, BBN considerations already
disfavour scenarios in which a 4th neutrino is populated.  If we avoid the
thermalisation of the sterile state (via the presence of a lepton asymmetry) it
may have two desirable consequences: In addition to eliminating the BBN
problems, the large scale structure mass limits may also be avoided.
For instance, if we have a 3+1 model in which the heaviest, isolated, mass 
state consists mostly of the sterile neutrino, the abundance of this heavy 
state (and hence its contribution to $\Omega_\nu$) will be small.

\section{Conclusions}
Large angle MSW transitions lead to neutrino flavour
equilibration in the early universe. This sets the strongest limit on the
universe's lepton number, because stringent constraints on the $\nu_e$ --
$\overline{\nu}_e$ asymmetry can now be applied to all three flavours.
The possibility of ``degenerate'' BBN is thus eliminated, thereby
removing a possible uncertainty in cosmological determinations of
neutrino mass.

Sterile neutrino are cosmologically disfavoured if they are brought into 
thermal equilibrium before BBN.  While all sterile neutrino models which can 
accommodate LSND suffer this problem, the constraints can be avoided by the
presence of a lepton asymmetry, which prevents thermalisation of the sterile, 
thus avoiding both BBN limits and large scale structure mass limits.

\newpage

{\bf Acknowledgments -- }  
N.F.B. was supported by Fermilab (operated by URA under
DOE contract DE-AC02-76CH03000) and by NASA grant NAG5-10842.

\end{document}